\documentclass{article}
\usepackage[utf8]{inputenc}
\usepackage{graphicx}
\usepackage{subfigure}
\usepackage{caption}
\usepackage{xcolor}
\usepackage{url}

\title{How SVC enables Distributed Caching in MEC?}
\author{Suvadip Batabyal\\Department of Computer Science and Information Systems, \\
BITS Pilani Hyderabad Campus, India.\\sbatabyal@hyderabad.bits-pilani.ac.in}

\begin{document}

\maketitle

FOR THE SPECIAL ISSUE ON SYSTEMS OPTIMIZATION

\section{Abstract}
With an ever increasing demand for the delivery of internet video service, the service providers are facing a huge challenge to deliver ultra-HD (2k/4k) video at sub-second latency. The multi-access edge computing (MEC) platform actually helps in achieving this objective by caching popular contents at the edge of cellular network. This not only reduces the delivery latency, but also the load and the cost of the backhaul links. However, MEC platforms are afflicted by constrained resources in terms of storage and processing capabilities; and centralized caching of contents may nullify the advantage of reduced latency by lowering the offloading probability. Distributed caching at the edge not only improves the offloading probability, but also dynamically adjusts the load distribution among the MEC servers. In this article, we propose an architecture for deployment of MEC platforms by exploiting the characteristics of a scalable video encoding technique. The layered video coding techniques, such as the scalable video coding (SVC), is used by the content providers to adjust to the network dynamics, by dynamically dropping packets in order to reduce latency. We show how an SVC video easily lends itself to distributed caching at the edge. Then we investigate the latency-storage trade-off by storing the video layers at different parts of the access networks.

\section{Introduction}
As per the Cisco annual report, nearly two-third of the global population will have access to the internet by 2023, with video services consuming the majority of the bandwidth provided by the mobile network operators and internet service providers. Globally, devices and connections are growing faster (10\% CAGR) than both population (1.0\% CAGR) and the internet users (6\% CAGR). This increase is putting a tremendous challenge on the service providers to meet the consumer's requirement for bandwidth, especially for the delivery of ultra-HD or 2k and 4K videos. Additionally, there is an increasing demand for videos having bit-rate (15-18Mbps for 4K videos) which is augmenting the challenge of best service delivery. With this fact, the mobile network operators (MNOs) and service providers are (i) deploying ultra-dense networks, for energy-efficient coverage of urban landscape, (ii) deploying multi-access edge computing (MEC) platforms, for low-latency delivery of multimedia contents and reduce the transport cost of the backhaul networks.

\subsection{Ultra-dense Networks}
Ultra-dense networks (UDNs)\cite{4} are realised through deployment of small cells, which are access points with small transmission power (and hence coverage). These access points cater to the service requirement of a significantly small percentage of population. UDNs facilitate reuse of frequency spectrum in an energy efficient manner and also improves the link quality (bits/hertz/unit area). The small cell access-points may be a fully functional base station (picocell or femtocell) or simply a remote radio head (RRH). The fully-functioning BS is capable
of performing all the functions of a macrocell with a lower
power in a smaller coverage area. The small cell BSs (SBSs) may be connected to the macrocell BS (MBS) using either a wireless or a wired backhaul technology. Although wired backhaul is capable of providing a higher data-rate, due to growing number of SBS, wireless backhaul is emerging to be more cost effective and scalable.

\subsection{Multi-access Edge Computing}
MEC is a natural development in the evolution of mobile base stations and the convergence of IT and telecommunications networking. Multi-access Edge Computing will enable new vertical business segments and services for consumers and enterprise customers. It allows the software applications to efficiently tap into local content and obtain real-time information about local-access network conditions. By deploying various services and caching contents at the network edge, mobile core networks are alleviated of further congestion and can efficiently serve local purposes. MEC uses case include computation offloading, distributed caching and content delivery, enhanced web performance, etc.

The MEC architecture\cite{1} typically consists of a host level and a system level manager. The MEC system level management includes the multi-access edge orchestrator (MEO) as its core component, which has an overview of the complete MEC system. The MEO has the visibility over the resources and capabilities of the entire mobile edge network
including a catalog of available applications. 

\section{Deployment Architecture}
\begin{figure}
    \centering
    \includegraphics[width=1.0\linewidth]{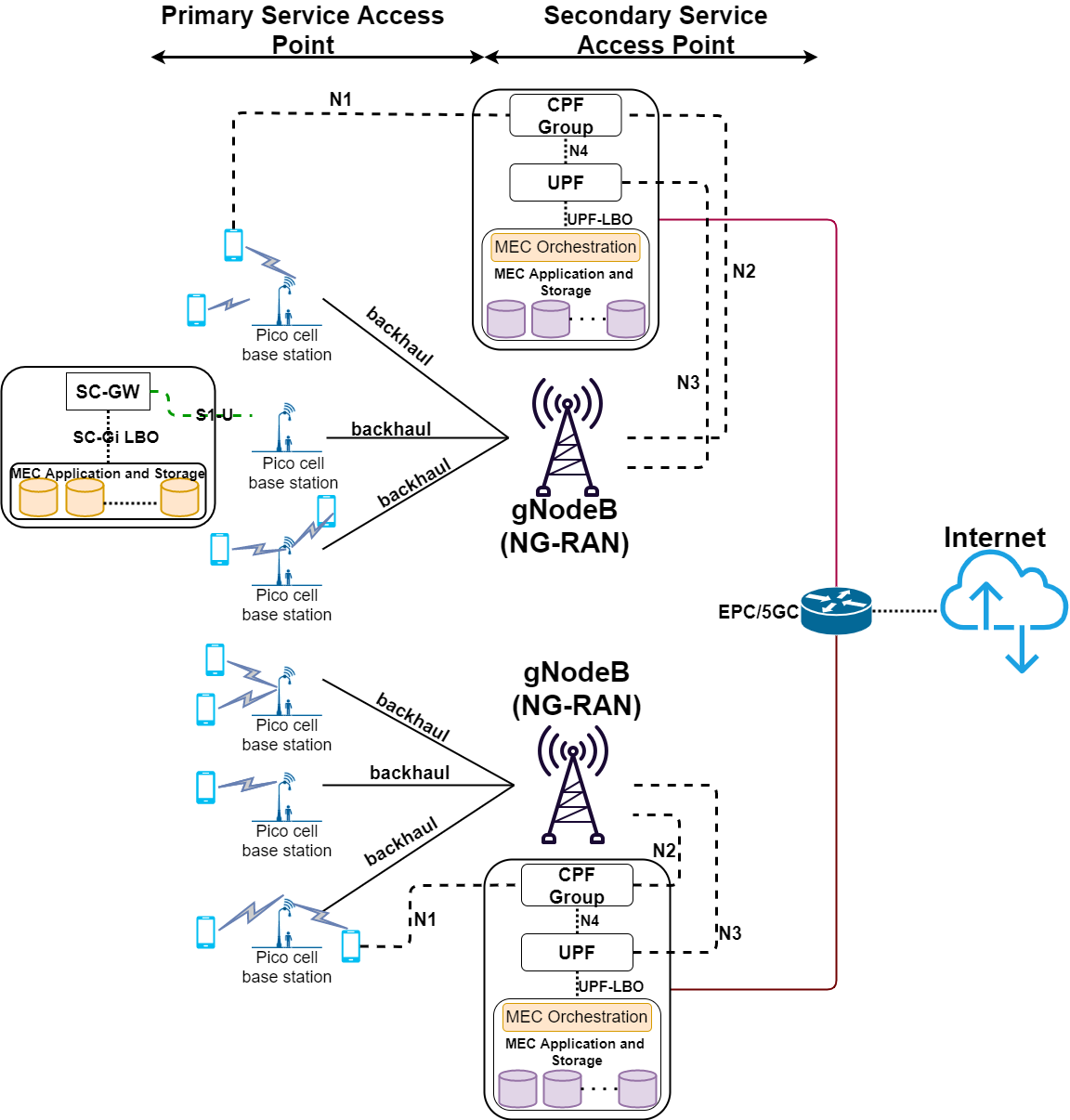}
    \caption{Architecture for SVC based distributed caching}
    \label{fig:1}
\end{figure}

Figure \ref{fig:1} shows the proposed deployment architecture. The architecture is divided into two slices (i) primary service access points (P-SAPs), and (ii) secondary service access points (S-SAPs). The P-SAPs consists of the small-cell base stations along with the MEC platform. The S-SAPs consists of the macro-cell base stations along with the MEC platform. The user equipments (UEs) are typically attached to the P-SAPs and gets serviced by it. The UE gets attached to the S-SAPs only under zero coverage by the P-SAPs. The P-SAPs consists of two main modules viz., the user plane functionality (UPF) and the MEC platform. The S-SAPs consists of three main modules viz., the user plane functionality (UPF), the control plane functionality (CPF), and the MEC platform. The UPF in P-SAPs consists of small cell serving gateway (SC-GW) \cite{2}. SC-GW is primarily responsible for routing of traffic to and from core network to the UEs. It can also take care of handover under mobility from one SBS to another. The SC-GW is attached to the MEC platform using SC service gateway interface with local breakout (SC-Gi LBO). 
It may be mentioned here that both the SC-Gi LBO and the MEC application may be hosted as VNFs (virtual network functions) in the same MEC platform. In S-SAPs, the CPF consists of various functionalities such as network exposure functions (NEF), network resource function (NRF), authentication server function (ASF), etc. The UPF is connected to the CPF, bearing the service gateway, through the N4 interface. Similar to the P-SAP, the UPF is connected to the MEC platform using the S-GW LBO. 

The MEC platform (in both P-SAPs and S-SAPs) primarily consists of two components viz., the MEC orchestrator (MEO), and MEC application and storage. The MEO is responsible for maintaining an overall view of the MEC system based on deployed MEC hosts, available resources, available MEC services, and topology, etc. The MEC application services may include computation offloading, content delivery, edge video caching, etc. The storage system stores the contents to be delivered to the UE in real-time during a service request.

We exploit the characteristics of layered coding technique for distributed caching\cite{6}\cite{7} in the proposed architecture. A layered video coding technique, such as the scalable video coding (SVC), consists of a base layer with multiple enhancement layers. The number of layers in a video (besides other factors) depends on the desirable video size (or downloadable segment size) for the application, and the level of scalability required. The enhancement (or higher) layers are encoded on the base layer (or layer-0). Moreover, each higher layer is encoded to its previous layers. The reception  of  the stream’s base layer is mandatory to play the video without a stall or a skip, albeit at a lower quality, and also to decode the enhancement layers. With the reception of enhancement layers, the video quality keeps improving. We propose two schemes for caching any video at the MEC platform. Case 1: where the base layer of a video is stored at the SBS MEC and the enhancement layers are stored in MBS MEC. Case 2: where the base layer is stored in the MBS MEC and the enhancement layers are stored in SBS MEC. The above caching decision is decided by the MEO that is hosted at the secondary service access point, and essentially in the gNodeB. The MEO has the visibility over the resources and capabilities of the entire mobile edge network including a catalog of available applications. The MEO (in the current context) decides whether to cache a certain video segment, based on the available resources and system performance. This decision is taken based on the availability of the storage, access profile of the video, and other video characteristics. Requests for access to video are first handled by the SBS. If available, the user is serviced; this is also notified to the MEO. Else, the MEO handles the requests and also makes a decision whether to forward the video the SBS for storage. The MEO runs daemon processes to track system parameters based on which such decisions are taken. We analyse the latency and storage capacity trade-off for the above schemes and determine the factors which affect the performance of both the schemes.

\section{Trade-off Analysis}
We perform a cache capacity to latency trade-off using a SVC video. The video has 20 GOPs where each GOP is of 2 seconds. The SVC is encoded at a constant bit-rate of 16Mbps, which is typically the standard \emph{average} bit-rate for a 2K video having a frame-rate of 30Hz. Each video has a base-layer and 4 enhancement layers. The enhancement layers contribute to the 20\% of the GOP size, the remaining 80\% being contributed by the base-layer. Therefore, given the settings, the size of a GOP is 32Mb, out of which 25Mb (approximately) is the size of the base-layer; and all the enhancement layers (taken together) makeup 7Mb. 

Next we assume that the SBS can provide a data rate (depending upon different configuration) varying between 100Mbps to 1000Mbps (although practically data-rates above 500Mbps are still a vision) and the MBS can provide a data-rate varying between 10Mbps to 100Mbps. We assume that a MBS has a storage capacity four times that of SBS. A similar assumption and analysis can be found in \cite{5}. It is to be mentioned here that there exists an asymptotic relation of storage size and download data rate between SBS and MBS. Therefore, any other experimental parameter that adhere to the asymptotic relation will not change the basic inferences.

\subsection{Caching Rules}
We assume two caching rules viz., \begin{enumerate}
    \item \textbf{Rule-1}: when the base-layer of a video is stored in the SBS cache and enhancement layers are stored in MBS. In this case, if the base-layer of a video is found in SBS, then the enhancement layer will be found in MBS. However, the vice-versa is not true. If the base-layer is not found in SBS, then the enhancement layer may or may not be there in the MBS, depending upon the popularity of the video.
    \item \textbf{Rule-2}: when the base-layer of a video is stored in MBS and enhancement layers are stored in SBS. In this case, if base-layer is found in MBS, then the enhancement layer will be found in SBS and vice-versa.
\end{enumerate}

Several cache replacement strategies have been proposed in this context \cite{3}. We assume a popularity based cache replacement strategy here.

\subsection{Results}

\begin{figure*}
    \centering
    \subfigure[]{\includegraphics[width=0.45\linewidth]{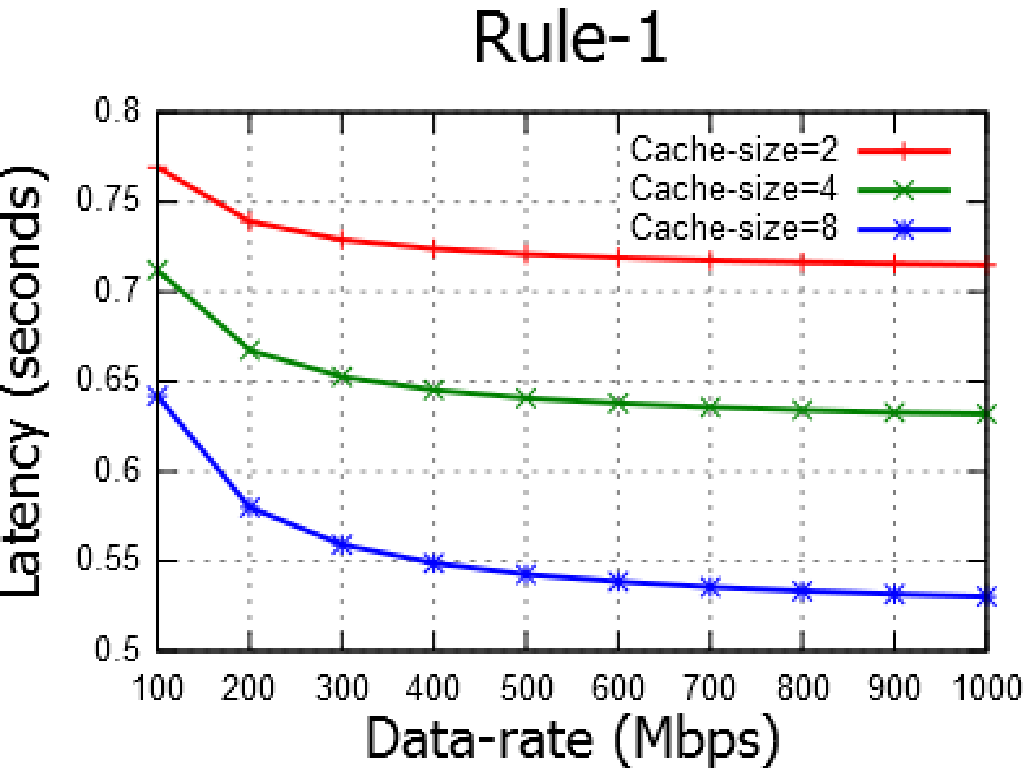}
    \label{fig:2a}}
    \subfigure[]{\includegraphics[width=0.45\linewidth]{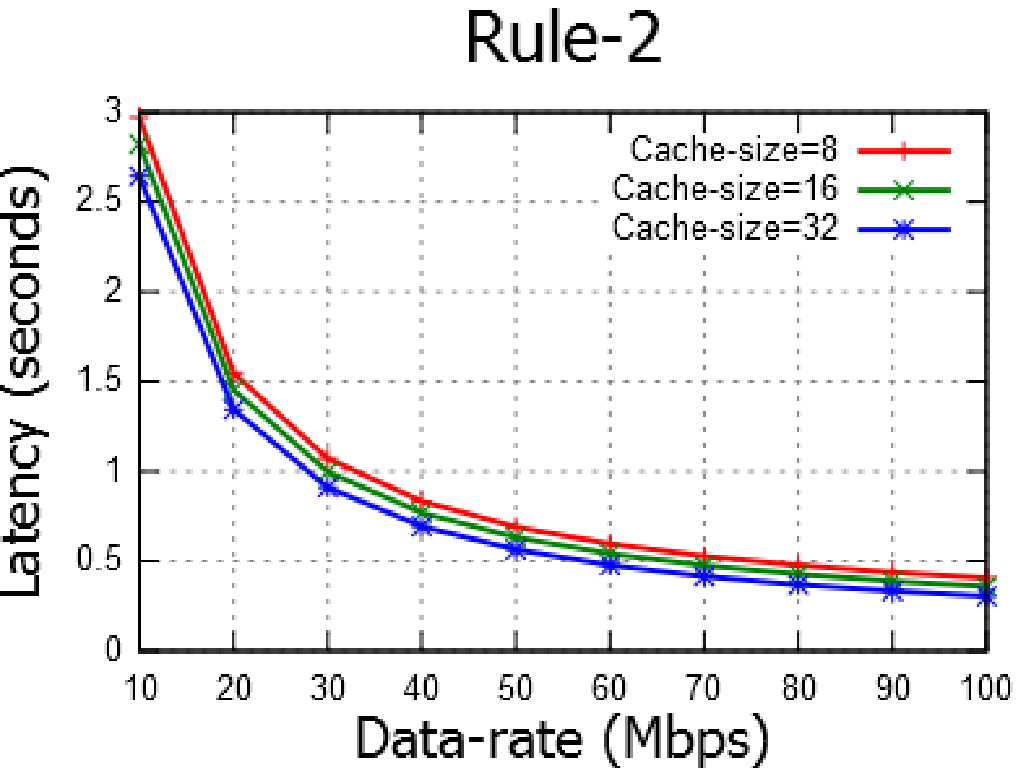}
    \label{fig:2b}}
    \caption{Graph of latency (seconds) against data-rate (Mbps) for (a) Rule-1 (b) Rule-2.}
\end{figure*}

Figure 2a and 2b shows the graph of latency against data-rates for 3 different cache sizes under rule-1 and rule-2, respectively. It can be observed that:
\begin{enumerate}
    \item Average latency under rule-1 is significantly smaller compared to rule-2. This is due to the higher data-rate in SBS. 
    \item The rate of decrease in latency is higher under rule-2 compared to rule-1 as data-rate is improved. This is because of asymptotically higher data rates in SBS compared to MBS, even for smaller cache hit-ratio.
    \item Latency is smaller in MBS compared to SBS when data rate is 100Mbps. This is because of the higher cache hit-ratio in MBS.
    \item Cache hit-ratio \emph{does not} play a significant role under asymptotically smaller data-rates viz., in case of rule-2. This is because of the bottleneck caused due to the lower transmission rate aided with the latency of the backhaul network.
\end{enumerate}

\section{Conclusion and future directions}
We proposed an architecture for distributed caching of an SVC video and performed a latency-storage trade-off. It is envisaged that higher storage capacity at MEC incurs higher cost and requires higher power supply. Therefore, SBS can support only small cache size compared to MBS. We find that SBS yields significantly improved latency under higher data rates. On the other hand MBS can yield similar latency under higher hit-ratio that can be obtained by higher cache size. In future, it would be interesting to investigate latency and cost trade-off, where cost may be calculated as the weighted sum of storage capacity and bandwidth.

\end{document}